\documentstyle{elsart}
\advance\textheight by \topskip

\begin{document}
\begin{frontmatter}
\begin{flushright}
NBI-HE-96-27\\
{\tt hep-th/9606099}
\end{flushright}
\title{{\bf 
Proof of universality of 
the Bessel kernel for\\ ~\,chiral matrix models 
in the microscopic limit
}}
\author{Shinsuke Nishigaki}
\address{Niels Bohr Institute, Blegdamsvej 17,
DK-2100 Copenhagen \O, Denmark}
\begin{abstract}
We prove the universality of
correlation functions of chiral complex matrix models
in the microscopic limit 
($N\to\infty,\ z\to 0,\ N z=$ fixed) 
which magnifies the crossover region
around the origin of the eigenvalue distribution.
The proof exploits the fact that the 
three-term difference equation for orthogonal polynomials
reduces into a universal second-order differential 
(Bessel) equation in the microscopic limit.

\end{abstract}
\end{frontmatter}
\footnotetext{e-mail address: {\tt nishigaki@nbi.dk}}

The concept of universality in the theory of 
random matrices,
or the independence of relevant quantities 
upon details of the potential function,
is crucial in its application
toward the level statistics of
disordered physical systems.
The universality of the macroscopic bulk two-point function
in the large-$N$ limit is discovered to hold \cite{AM} 
as a direct consequence of 
the linear functional relationship
(Cauchy inversion) 
between the potential $V(M)$ and the large-$N$ 
spectral density
$\rho(z)=\lim_{N\to\infty}\left\langle 
1/N\,{\rm tr}\,\delta(z-M)\right\rangle$
\begin{equation}
\frac{V'(z)}{2}=
-\!\!\!\!\!\!\int_{-a}^a \frac{\d w}{z-w}\, \rho(w)
\ \ \Leftrightarrow \ \ 
\rho(z)=
-\frac{1}{\pi^2}
-\!\!\!\!\!\!\int_{-a}^a \frac{\d w}{z-w}\, 
\sqrt{{a^2-z^2}\over{a^2-w^2}}
\frac{V'(w)}{2}
\label{bipz}
\end{equation}
\cite{BIPZ},
and that $\rho(z,w)\sim\delta \rho(z)/\delta V(w)$.
An alternative proof using orthogonal polynomials is
found in ref.\cite{BZ}.

Another class of universality of a different origin, 
termed microscopic,
has been anticipated whenever a quantity in concern
is governed by the microscopic repulsion (anti-crossing) 
between energy levels 
dominantly enough to surpass the effect of
a slowly varying potential.
This type of (conjectured) universalities includes
the appearance of the sine kernel for the microscopic
correlators in the bulk \cite{Wig,Gor}\footnotemark[1]
(proved in ref.\cite{KPW}),
of the Airy kernel in the vicinity of a `soft edge' 
($\rho(z)\sim\sqrt{a-z}$) \cite{Moo}\footnotemark[1]
(proved in ref.\cite{BB}), 
and of the Bessel kernel of a `hard edge' 
($\rho(z)\sim 1/\sqrt{a-z}$) \cite{NS,For}\footnotemark[1].
The last and open problem has attracted 
considerable attention from
condensed matter physics on spin-impurity 
scattering \cite{HZ}
as well as from high energy physics on 
QCD chiral symmetry breaking
\cite{V93}, and will be the central subject of this article.

The problem can equivalently be formulated in terms of 
chiral complex matrix models as follows:
Consider a matrix integral with a generic potential
\begin{equation}
Z=\int \d^{2N^2}M \, \exp 
\left\{ -\frac{N}{2} {\rm tr}\, V(M^2) \right\},\ \ \ 
V(M^2)=\sum_{k\geq 1} \frac{g_k}{k}M^{2k}
\label{partition} 
\end{equation}
where $M$ stands for a $2N\times 2N$ block hermitian matrix
whose non-zero components are $N\times N$ complex matrices
on the off-diagonals,
\begin{equation}
M=
\left(
\begin{array}{cc}
0&C^\dagger\\
C&0
\end{array}\right).
\end{equation}
A `chiral' complex matrix model (or chiral unitary ensemble)
is so called because of 
the invariance under the transformation
\begin{equation}
C\mapsto U\,C\,V^\dagger, \ \ U, V \in {\rm U}(N).
\label{chi-inv}
\end{equation}
Since $M$ anticommutes with 
$\gamma_5=
{\scriptsize
\left(
\begin{array}{cc}
I&0\\
0&-I
\end{array}\right)}
$, 
each of its eigenvalues $z$ always accompanies 
its mirror image $-z$
in the spectrum. 
The repulsion between these pairs is expected to
yield a region in the vicinity of
 the origin where eigenvalues 
avoid populating.
Magnify this region by measuring the correlation functions
in the unit of an average level spacing $\sim 1/N$, i.~e.\ 
by substituting $z=\zeta/N$. 
Are all the correlators independent of
the potential
in the limit $N\to\infty$, when $\zeta$ is kept fixed?

The answer has already been conjectured affirmative 
on several grounds \cite{TW,BHZ}. 
First, one na\"{\i}vely expects that, 
in the vicinity of the origin, 
the potential can essentially be regarded as a constant.
Therefore it would not affect the microscopic correlation
due to the level repulsion, 
except via the average level spacing 
used as a unit, which is determined by global balancing. 
Ref.\cite{TW} noticed that one and
the same Bessel kernel emerges from two ensembles with
distinct potentials, Gaussian $V=M^2$ 
and an infinite well\footnotemark[1].
\footnotetext[1]{These results for simple potentials
are corollaries to the well-known asymptotic behaviour 
of classical orthogonal polynomials \cite{Sze}.}
This suggests that the Bessel kernel holds 
for a generic potential.
Ref.\cite{BHZ} has calculated the one-point function
for $V=M^2+g\,M^4$ up to $O(g^1)$, which again 
supports the conjectured universality.
We shall give a rigorous proof of this conjecture
by exploiting the fact that the three-term difference 
equations for orthogonal polynomials,
characteristic of one-matrix models, 
universally reduces the Bessel equation 
in the above mentioned microscopic limit.\\

The partition function for a chiral complex matrix model 
(\ref{partition}) is expressible in terms of the 
component matrices
as well as of the eigenvalues after integration over 
the angular coordinates
$(U,V)\in {\rm U}(N)\times{\rm U}(N)/{\rm U}(1)^N$,
\begin{eqnarray}
Z&=&\int \d^{N^2}C^\dagger\, \d^{N^2} C\, 
\e^{-N{\rm tr} V(C^\dagger C) }
\nonumber \\
&=&
\int_{-\infty}^\infty \prod_{i=1}^N 
\left( \d z^2_i\, \e^{-N V(z_i^2) }\right)
\prod_{i<j}\left|z_i^2-z_j^2 \right|^2 =
\int_0^\infty \prod_{i=1}^N \left( \d\lambda_i\, 
\e^{-N V(\lambda_i) }\right)
\prod_{i<j}\left|\lambda_i-\lambda_j\right|^2 .
\end{eqnarray}
The above expression can be interpreted as a 
{\em positive definite}
hermitian matrix model in $H=C^\dagger C$ whose 
eigenvalues are $\lambda_1,\cdots,\lambda_N\geq 0$.
In other words, the problem reduces to finding 
a set of orthogonal polynomials $P_n(\lambda)$
over the semi-infinite interval $[0,\infty)$,
\begin{equation}
\int_0^\infty \d\lambda\, \e^{-N V(\lambda) }
 P_n(\lambda)\,P_m(\lambda)=
h_n\, \delta_{nm}.
\label{orthcomp}
\end{equation}
We normalise them such that $P_n(0)=1$ 
for later convenience,
\begin{equation}
P_n(\lambda)=1+\cdots+{p_n}{\lambda^n}.
\label{normal}
\end{equation}
Here it is assumed possible to choose this normalisation,
which will prove equivalent to the ansatz that
the origin be included in the support of
the large-$N$ spectral density of $M$.
Then the recursion relation for these $P_n$'s reads
\begin{eqnarray}
\lambda\, P_n(\lambda)&=&
-q_{n} P_{n+1}(\lambda)+s_n P_n(\lambda)
-q_{n-1}\frac{h_{n}}{h_{n-1}} P_{n-1}(\lambda)
\ \ \ \ \ \left( q_n\equiv-\frac{p_{n}}{p_{n+1}}\right)
\label{3t-rec}\\
&\equiv& \sum_m \hat{\lambda}_{nm}\,P_m(\lambda)\nonumber.
\end{eqnarray}
The sets of unknowns $\{h_n\}$, $\{q_n\}$,  $\{s_n\}$ are 
iteratively determined by \cite{Mor}
\begin{eqnarray}
&&1=
-\int_0^\infty 
\d\lambda \frac{\d}{\d\lambda} \left\{ 
\e^{-N\,V(\lambda)}P_n(\lambda)\,P_{n}(\lambda) \right\}=
N\,V'(\hat{\lambda})_{nn}\, h_n ,
\label{vnn}\\
&&1=
-\int_0^\infty 
\d\lambda \frac{\d}{\d\lambda} \left\{ 
\e^{-N\,V(\lambda)}P_{n}(\lambda)\,
P_{n-1}(\lambda) \right\}=
\left( N\,V'(\hat{\lambda})_{nn-1}
+  \frac{n}{q_{n-1}} \right) h_{n-1}
\label{vnn-1}
\end{eqnarray}
and eq.(\ref{3t-rec}) at $\lambda=0$ ,
\begin{equation}
0=-q_{n}+s_n -q_{n-1}\frac{h_{n}}{h_{n-1}} .
\label{sn-eq}
\end{equation}
We can immediately eliminate $s_n$'s 
using (\ref{sn-eq}) to get
\begin{equation}
\lambda\,P_n(\lambda)=
-q_{n}\left\{ P_{n+1}(\lambda)- P_n(\lambda)
-\frac{h_{n}}{q_n}\frac{q_{n-1}}{h_{n-1}} 
\left( P_{n}(\lambda)-P_{n-1}(\lambda) 
\right) \right\}.
\label{3t-recII}
\end{equation}
\setcounter{footnote}{1}

In the following we need to know 
the asymptotic behaviour of 
$q_n$ and $h_n$ for 
\begin{equation}
n,\ N\to \infty\ \  
\mbox{while}\ \ \frac{n}{N}=t\ \mbox{is kept fixed}.
\label{lim}
\end{equation}
Eqs.(\ref{vnn}), (\ref{vnn-1}) and (\ref{3t-recII})
tell us that
they should behave as\footnote{$q_n$ and $h_n$
converge to smooth functions when the eigenvalues
are supported on a single interval.}
\begin{equation}
q_n=q(\frac{n}{N})
+ \mbox{higher orders in }\frac{1}{n}, \ \ \ 
N\,h_n=h(\frac{n}{N})
+ \mbox{higher orders in }\frac{1}{n}. 
\end{equation}
Then 
the matrix $\hat{\lambda}$ and its powers are
approximated to be
\begin{eqnarray}
&&\hat{\lambda}_{nm}= q(\frac{n}{N})
\left(
-\delta_{n\,m-1}+2\delta_{n m}-\delta_{n\,m+1}
\right), \nonumber\\
&&(\hat{\lambda}^k)_{nm}= q(\frac{n}{N})^k
\sum_{\ell=-k}^k (-)^\ell \left( {2k \atop k+\ell} \right)
\delta_{n\,m+\ell}
\end{eqnarray}
so that eqs.(\ref{vnn}) and (\ref{vnn-1}) read
\begin{eqnarray}
\sum_{k} g_k \left({2k-2\atop k-1}\right) 
q(t)^{k-1}
&=&
\frac{1}{h(t)} ,
\label{vNN}\\
-\sum_{k} g_k \left({2k-2 \atop k}\right) 
q(t)^{k-1}
&=&
\frac{1}{h(t)}
-\frac{t}{q(t)} .
\label{vNN-1}
\end{eqnarray}
By eliminating $h(t)$ out of the above two, 
we obtain
an algebraic equation for $q(t)$,
\begin{equation}
\frac12 \sum_{k} g_k \left({2k \atop k}\right) 
q(t)^{k} =t.
\label{qn}
\end{equation}
Eqs.(\ref{vNN}) and (\ref{qn}) 
imply a universal relationship
among total derivatives,
\begin{equation}
\d t=
2q\,\d\left( \frac1h \right) + \frac{1}{h} \d q
=2\sqrt{q}\, \d \left( \frac{\sqrt{q}}{h} \right).
\label{dt}
\end{equation}

Next we expand the rhs of the 
recursion equation (\ref{3t-recII})
in terms of $1/n$ in the limit (\ref{lim}),
\begin{equation}
\lambda\,P(n,N,\lambda)=
-\frac{q(t)}{N^2} 
\left\{ \frac{\d^2}{\d t^2} +
\frac{h}{q} \left(\frac{\d}{\d t} \frac{q}{h} \right)
\frac{\d}{\d t} 
\right\} P(n,N,\lambda)
+ \mbox{higher orders in }\frac{1}{n}
\end{equation}
where the argument $N$ in 
$P(n, N, \lambda)\equiv P_n(\lambda)$ is to indicate 
explicitly the dependency
via the coefficient in front of the potential.
It equivalently reads
(subleading terms suppressed)
\begin{equation}
\left(
h(t) \frac{\d}{\d t}  \frac{q(t)}{h(t)} \frac{\d}{\d t}
+N^2 \lambda
\right)
 P(n,N,\lambda) 
=0
\label{diff-eq}
\end{equation}
telling us that that the arguments of $P$ appear
only in the combinations $t=n/N$ and $x=N^2\lambda$
in the limit (\ref{lim}).
The rescaled eigenvalue coordinate $x$ 
is to be fixed finite hereafter,
and is regarded as a parameter in the ordinary
differential equation in $t$.
Performing the change of variable
$
t \mapsto u(t)\equiv \sqrt{q(t)}/h(t),
$
using the relationship (\ref{dt})
and neglecting higher order terms in $1/n$,
eq.(\ref{diff-eq}) 
reduces to the Bessel equation of zeroth order,
\begin{equation}
\left(
\frac1u \frac{\d}{\d u}  u \frac{\d}{\d u} +4x
\right)
P(u,x)=0.
\end{equation}
The general solution to it is a linear combination of
Bessel and Neumann functions
\begin{equation}
P(u,x)=
c (x) \,J_0\left( 2u\sqrt{x} \right) +
c'(x) \,Y_0\left( 2u\sqrt{x} \right).
\label{Bes}
\end{equation}
The integration constants (functions in $x$) are
completely fixed by 
the boundary condition at $t=n/N=0$ ($u(0)=0$),
\begin{equation}
P(0, x)=P_0(\lambda)=1 ,
\end{equation}
to be $c(x)=1,\ c'(x)=0$.
By substituting $t=1$
into (\ref{Bes}) 
and its $t$-derivative,
we establish the following lemma for the asymptotic
behaviour of generic orthogonal polynomials over
the semi-infinite range $[0, \infty)$ which are
normalisable to $P_n(0)=1$:
\begin{eqnarray}
&&
\lim_{N\to\infty}
P_N( \frac{x}{N^2})=
J_0\left(2u(1)\sqrt{x}\right) ,
\label{J0} \\
&&
\lim_{N\to\infty}
N\left( P_N( \frac{x}{N^2}) 
-P_{N-1}( \frac{x}{N^2})\right) 
=-\sqrt{x\over q(1)}\,
J_1\left(2u(1)\sqrt{x}\right) .
\label{J1}
\end{eqnarray}
The parameters $u(1)$ and $q(1)$, 
through which the dependence
upon the potential enters the asymptotic
form of the orthogonal polynomial, have the following
simple meaning.
Comparison of
eqs.(\ref{vNN}) and (\ref{qn}) at $t=1$
with an explicit expression for the rhs of eq.(\ref{bipz})
\begin{eqnarray}
&&\rho(z)=\frac{\sqrt{a^2-z^2}}{2\pi} 
\sum_{k} g_k \sum_{n=0}^{k-1}
\left( {2n \atop n} \right) \left({a^2 \over 4}\right)^n
z^{2k-2n-2},\\
&& \frac12 \sum_{k} g_k \left({2k \atop k}\right) 
\left( {a^2 \over 4}\right)^{k}=1
\end{eqnarray}
enables us to relate the parameters with
the edge of the support of the large-$N$
spectral density $\rho(z)$ 
of the chiral complex 
matrix and its value at the origin as
\begin{equation}
a=2\sqrt{q(1)},
\ \ 
\rho(z=0)=\frac{u(1)}{\pi}.
\label{rho0}
\end{equation}
It is easy to check that the critical condition 
$\rho(0)=0$
for the 1-/2-cut transition \cite{Shi}
is equivalent to $P_N(0)=0$ 
for the conventional, monically normalised polynomials
$P_n(\lambda)=\lambda^n+\cdots$.
Thus, under the assumption 
that the normalisation (\ref{normal}) is possible,
the constants $h(1)$ and $q(1)$
are determined positive,
due to the identification (\ref{rho0})
valid for $\rho(z)$ supported on a single interval.
The positivity of the norm $h$ 
is necessary for consistency,
whereas $q>0$ signifies that the coefficients in 
$P_n(\lambda)$ are alternating and eventually
causes an oscillatory microscopic spectral density 
as it should.

Now we recall the expression for 
the integration kernel $K_N(\lambda,\mu)$ 
associated with the eigenvalue problem
for the positive definite hermitian
matrix $H=C^\dagger C$,
\begin{eqnarray}
K_N(\lambda,\mu)&=&
\e^{-\frac{N}{2}(V(\lambda)+V(\mu))} \,
\frac1N \sum_{i=0}^{N-1} 
\frac{P_i(\lambda)P_i(\mu)}{h_i} \nonumber \\
&=&
\e^{-\frac{N}{2}(V(\lambda)+V(\mu))} \,
\frac{-q_{N-1}}{N\,h_{N-1}}
\frac{P_N(\lambda)P_{N-1}(\mu)-
P_{N-1}(\lambda)P_{N}(\mu)}{\lambda-\mu}.
\end{eqnarray}
Here use is made of the Christoffel-Darboux identity.
Plugging in the lemmata (\ref{J0}) and (\ref{J1}),
we obtain a universal form of the
kernel (called the Bessel kernel)
in the microscopic limit
\begin{eqnarray}
&&\lim_{N\to\infty} \frac1N\, 
K_N(\frac{x}{N^2}, \frac{y}{N^2})=
\nonumber\\
&&-2u(1) 
\frac{
J_0 ( 2u(1)\sqrt{x} ) 
\sqrt{y} J_1 ( 2u(1)\sqrt{y} )- 
\sqrt{x} J_1 ( 2u(1)\sqrt{x} ) 
J_0 ( 2u(1)\sqrt{y} )
}{x-y}. 
\end{eqnarray}
The $s$-point correlation function $\sigma_N$ ($\rho_N$)
of eigenvalues of $H$ ($M$)
is represented in terms of the kernel as
\begin{eqnarray}
&&\sigma_N(\lambda_1,\cdots,\lambda_s)=
\left\langle \prod_{a=1}^s \frac{1}{N} 
{\rm tr}\, \delta (\lambda_a-H)\right\rangle=
\det_{1\leq a,b \leq s} K_N (\lambda_a,\lambda_b)\\
&&\rho_N(z_1,\cdots,z_s)=
\left\langle \prod_{a=1}^s \frac{1}{2N} 
{\rm tr}\, \delta (z_a-M)\right\rangle=
| z_1 | \cdots | z_s |\, 
\sigma_N(z_1^2,\cdots,z_s^2),
\end{eqnarray}
respectively.
Therefore all the formulae for their 
microscopic limits
\begin{eqnarray}
&&\varsigma(x_1,\cdots,x_s)\equiv
\lim_{N\to\infty}\frac{1}{N^s}
\sigma_N (\frac{x_1}{N^2},\cdots,\frac{x_s}{N^2})\\
&&\varrho(\zeta_1,\cdots,\zeta_s)\equiv
\lim_{N\to\infty}
\rho_N(\frac{\zeta_1}{N},\cdots,\frac{\zeta_s}{N})=
| \zeta_1 | \cdots | \zeta_s |\, 
\varsigma(\zeta_1^2,\cdots,\zeta_s^2),
\end{eqnarray}
previously calculated for the 
Laguerre (in the $H$-picture) or 
chiral Gaussian (in the $M$-picture) 
unitary ensemble, hold universally.
Namely, 
the spectral density of the chiral complex matrix
model
\begin{equation}
\rho_N(z)=\left\langle \frac{1}{2N}
\, {\rm tr}\, \delta( z- M
)\right\rangle 
=| z | K_N (z^2, z^2)
\end{equation}
universally takes the form
\begin{equation}
\varrho (\zeta)=
\left( \pi \rho(0) \right)^2
|\zeta|
\left(
J_0^2(2\pi\rho(0)  \zeta) +
J_1^2(2\pi\rho(0)  \zeta)
\right)
\end{equation}
in the microscopic limit.
It enjoys the matching condition
between the micro- and macroscopic (large-$N$) 
spectral densities,
\begin{equation}
\lim_{\zeta\to\infty}\varrho(\zeta)=
\rho(0) .
\end{equation}
\vspace{5mm}

In this article we have exhibited a proof of universality
of the correlation functions of chiral complex 
matrix models in the microscopic limit. 
Consequently all dependencies of correlators
upon the potential
appears only through a single and local (at $z=0$) 
parameter $\rho(0)$ as anticipated.
The universality holds as long as $\rho(0)>0$,
i.e.~the large-$N$ spectral density is supported on
a single interval.

We have extended this strategy 
of taking the continuum limit of recursion equations,
commonly used in the double-scaling calculations,
to show microscopic universalities in
a variety of matrix models 
which have been argued to be of relevance
in QCD \cite{V94}.
Details of these generalisations will appear
elsewhere \cite{ADMN}.

\begin{ack}
I thank
P.~H.~Damgaard for many inspiring discussions.
The work of SN is supported by 
JSPS Postdoctoral Fellowships for Research Abroad.
\end{ack}


\begin{thebibliography}{99}
\bibitem{AM}
               J.~Ambj{\o}rn and Y.~M.~Makeenko,
               {\it Mod.~Phys.~Lett.} {\bf A5} (1990) 1753.
\bibitem{BIPZ} 
               E.~Br\'ezin, C.~Itzykson, 
               G.~Parisi, and J.~-B.~Zuber,
               {\it Commun.~Math.~Phys.} {\bf 59} (1978) 35.
\bibitem{BZ} 
               E.~Br\'ezin and A.~Zee, 
               {\it Nucl.~Phys.} {\bf B402} (1993) 613.
\bibitem{Wig}    
               E.~P.~Wigner, in: 
               {\it Statistical Properties of Spectra:
               Fluctuations}, ed.~C.~E.~Porter
               (Academic Press, New York, 1965), p.~446.
\bibitem{Gor}
               M.~Gaudin,
               {\it Nucl.~Phys.} {\bf 25} (1961) 447.
\bibitem{KPW}
               R.~D.~Kamien, H.~D.~Politzer and M.~B.~Wise,
               {\it Phys.~Rev.~Lett.} {\bf 60} (1988) 1995.
\bibitem{Moo}  
               G.~Moore,
               {\it Prog.~Theor.~Phys.~Suppl.} {\bf 102}
               (1990) 255.
\bibitem{BB}   
               M.~J.~Bowick and E.~Br\'ezin,
               {\it Phys.~Lett.} {\bf B268} (1991) 21.  
\bibitem{NS}
               T.~Nagao and K.~Slevin, 
               {\it J.~Math.~Phys.} {\bf 34} (1993) 2075.
\bibitem{For}
               P.~J.~Forrester,
               {\it Nucl.~Phys.} {\bf B402} (1993) 709.
\bibitem{HZ}                      
               S.~Hikami and A.~Zee,  
               {\it Nucl. Phys.} {\bf B446} (1995) 337.
\bibitem{V93}
               J.~J.~M.~Verbaarschot,
               {\it Acta.~Phys.~Pol.} {\bf B25} (1994) 133.
\bibitem{TW}
               C.~A.~Tracy and H.~Widom,
              {\it Commun.~Math.~Phys.} {\bf 161} (1994) 289.
\bibitem{BHZ}
               E.~Br\'ezin, S.~Hikami and A.~Zee,
               {\it Nucl.~Phys.} {\bf B464} (1996) 411.
\bibitem{Sze}  G.~Szeg\H{o}, 
               {\it Orthogonal Polynomials}  
               (Am.~Math.~Soc., Providence, 1939), Chap.VIII.
\bibitem{Mor}  
               T.~R.~Morris, 
               {\it Nucl.~Phys.} {\bf B356} (1991) 703.
\bibitem{Shi} 
               Y.~Shimamune, 
               {\it Phys.~Lett.} {\bf B108} (1982) 407.
\bibitem{V94}
               J.~J.~M.~Verbaarschot,
               {\it Nucl.~Phys.} {\bf B426} (1994) 559;\\
               J.~J.~M.~Verbaarschot and I.~Zahed,
               {\it Phys.~Rev.~Lett.} {\bf 73} (1994) 2288.
\bibitem{ADMN}  
               G.~Akemann, P.~H.~Damgaard, U.~Magnea and
               S.~Nishigaki,
               NBI preprint in preparation.
\end{thebibliography}
\end{document}